\documentclass[letterpaper,10pt]{article}
\usepackage{osameet2}
\usepackage{subcaption}
\usepackage{epstopdf}
\usepackage{amsmath,amssymb}
\usepackage{braket}
\usepackage[dvips,colorlinks=true,bookmarks=false,citecolor=blue,urlcolor=blue]{hyperref} 

\begin{document}

\title{Localization and delocalization dynamics of photons in linearly coupled cavity arrays}

\author{Nilakantha Meher and S. Sivakumar$^*$}
\address{Material Physics Division\\Indira Gandhi Centre for Atomic Research, Homi Bhabha National Institute, Kalpakkam, TamilNadu, India-603102}
\email{$^*$siva@igcar.gov.in}

\begin{abstract}
Localization and delocalization of two photons in an array of cavities are examined.  Role of   entanglement and relative phase of the initial state in  the occurrence of localization or delocalization during time evolution is elucidated.   
\end{abstract}


\section{Introduction}

    Waves and particles are treated differently in classical physics.  A particle is a point object described in terms of phase space points.   Waves, on the other hand,  are spatially extended, though  not necessarily over large domains.  In quantum physics, a particle is described by a suitable wave function which is vector in an appropriate Hilbert space.  This description implies that it is meaningful to talk about well-localized wave function for a particle.   If we consider the electromagnetic field, quantum theory describes it in terms of excitations in the electromagnetic vacuum,   for instance, photons corresponding to the field supported in a cavity.  If an array of coupled cavities is considered, it is possible to speak of photons being in one of the cavities or shared by more than one cavity. Hence, photon localization in a cavity is a physically meaningful situation\cite{Schmidt,Miran, Nil}.   Possibility of localization in two cavities has been studied with linear and nonlinear couplings\cite{ Sara, Nil}.  Presently,  we study  the localization and delocalization of two photons in  an array of $N$ linearly coupled cavities. In this case, two photon localization(TPL) corresponds to having  two photons in any one of the cavities on measurement.  If two photons are in two different cavities, it is two photon delocalization (TPD).

\section{Localization and delocalization in cavity arrays}

      We consider a chain of $N$ independent, linearly coupled cavities to discuss TPD and TPL.  
The Hamiltonian for the system is $(\hbar=1)$,
\begin{equation}\label{systemH}
H=\omega \sum\limits_{j=1}^N a_j^\dagger a_j +J \sum\limits_{j=1}^{N-1}(a_j^{\dagger} a_{j+1}+H.c).
\end{equation}
Here $a_j$ and $a_j^\dagger$ are the annihilation and creation operators corresponding to the field mode in $j$th cavity.   The photon 
hopping strength between two adjacent cavities is $J$. Using a suitable normal mode analysis, the time evolved operator for the $j$th mode is 
\begin{equation}\label{timeevolution}
a_j(t)=\sum_{l} G_{jl}(t) a_l(0),
\end{equation}
where $G_{jl}(t)=\sum_{k=1}^N e^{-i\Omega_k t} \tilde{S}(j,k) \tilde{S}(l,k)$, $\Omega_k=\omega+2J\cos\left(\frac{\pi k}{N+1}\right)$ and $\tilde{S}(j,k)=\sqrt{\frac{2}{N+1}}\sin(\frac{j\pi k}{N+1})$.   A similar expression can be derived for the creation operator $a_j^\dagger$. With these results, the dynamics of any physical quantity expressible in terms of the creation and annihilation operators can be determined.  Coupling, linear or otherwise, among the cavities leads to transport of  photons from one cavity to another in the array.  While it is a necessary condition for photon transport, realization of TPL or TPD is dependent on the initial state as well.\\  

In order to investigate the role of entanglement and phase in the initial state on the localization-
delocalization phenomenon, we consider states of the form
$\ket{\psi}=\cos\theta\ket{2}_r\ket{0}_s+e^{i\phi}\sin\theta\ket{0}_r\ket{2}_s$. 
The notation $\ket{p}_r\ket{q}_s$ stands for $p$ photons in $r$-cavity and $q$ photons in $s$-cavity and no photons in the other cavities of the array. Here $r$ and $s$ runs from 1 to $N$. 
Hence, the probabilities of detecting two photons in the $r$ and $s$ cavities are $\cos^2\theta$ and $\sin^2\theta$ respectively.   This is a two photon localized state according to our definition.  The relative phase $\phi$ does not influence the measurement outcomes.

As our focus is on localization and delocalization in the context of two photons, it is prudent to calculate joint probability $P_{mn}$ of coincidence detection of two photons at time $t$ in the cavities $m$ and $n$ in order to quantify TPL and TPD.  It is defined as  
\begin{equation}
P_{m,n}(t)=\frac{\langle a_n^{\dagger}(t)a_m^{\dagger}(t)a_m(t)a_n(t)\rangle}{\langle a_n^\dagger(t) a_n(t)\rangle\langle a_m^\dagger(t) a_m(t)\rangle}.
\end{equation}
 Here $\langle..\rangle$ refers to expectation value in the initial state and the time-developed operators are defined in Eq. \ref{timeevolution}. During the time evolution of $\ket{\psi}$ under $H$,
\begin{equation}
P_{m,n}(t)=\frac{1}{2}\frac{|\cos\theta G_{mr}(t)G_{nr}(t)+e^{i\phi}\sin\theta G_{ms}(t)G_{ns}(t)|^2}{(\cos^2\theta|G_{nr}|^2+\sin^2\theta|G_{ns}|^2)(\cos^2\theta|G_{mr}|^2+\sin^2\theta|G_{ms}|^2)}.
\end{equation}
The diagonal elements of this correlation matrix give the probability of localization.
If $n=m$, the expression gives the two photon detection probability which measures the degree of TPL.  Then the degree of TPD is \\
\begin{equation}\label{degreedeloc}
S=1-\sum_{n=1}^N P_{n,n}(t),
\end{equation}
which is essentially the probability of detecting photons in two different cavities. 
If $S=0$, it is TPL state and $S=1$ corresponds to TPD state.

\begin{figure}[h!]
\centering
\includegraphics[scale=0.3]{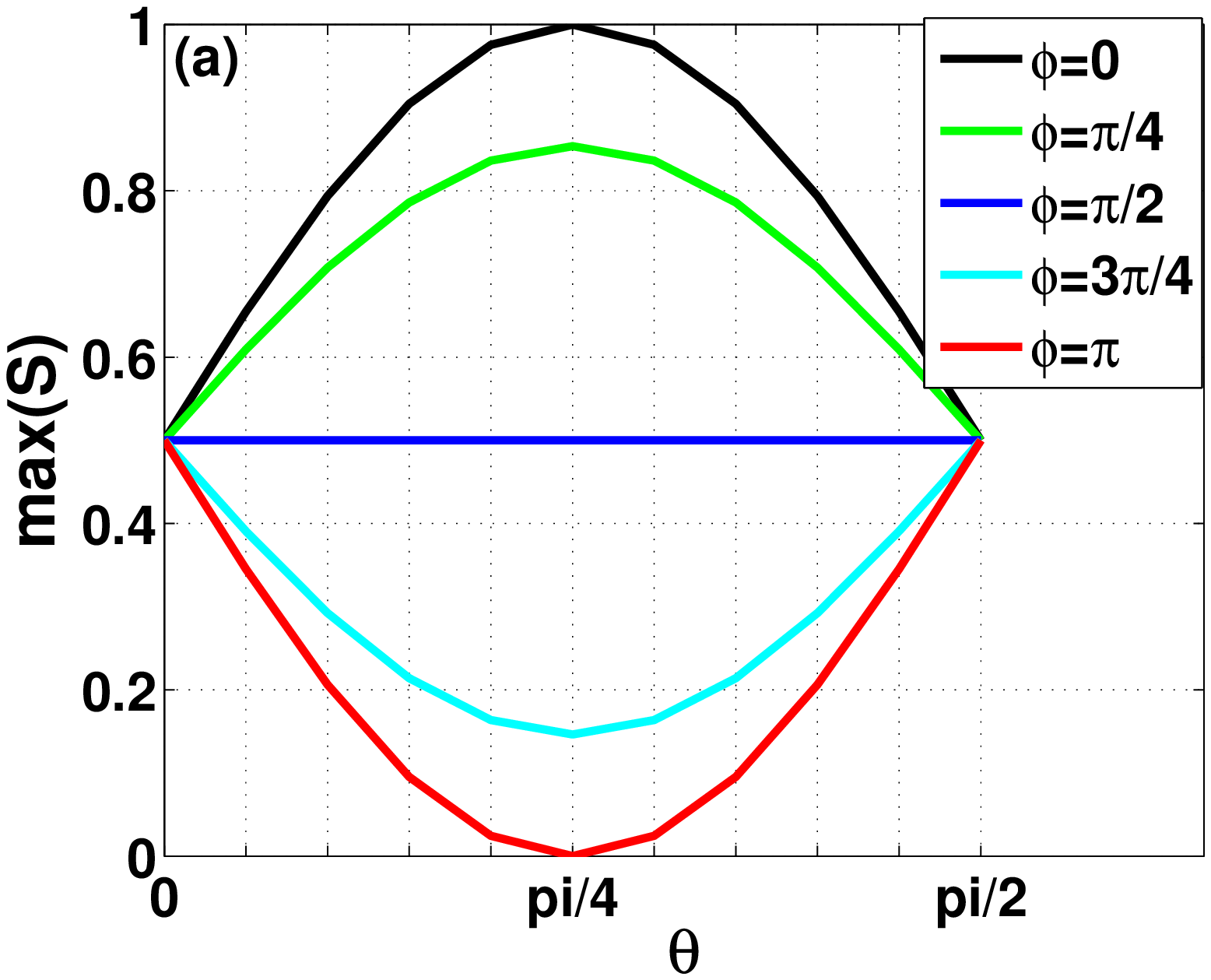}\includegraphics[scale=0.3]{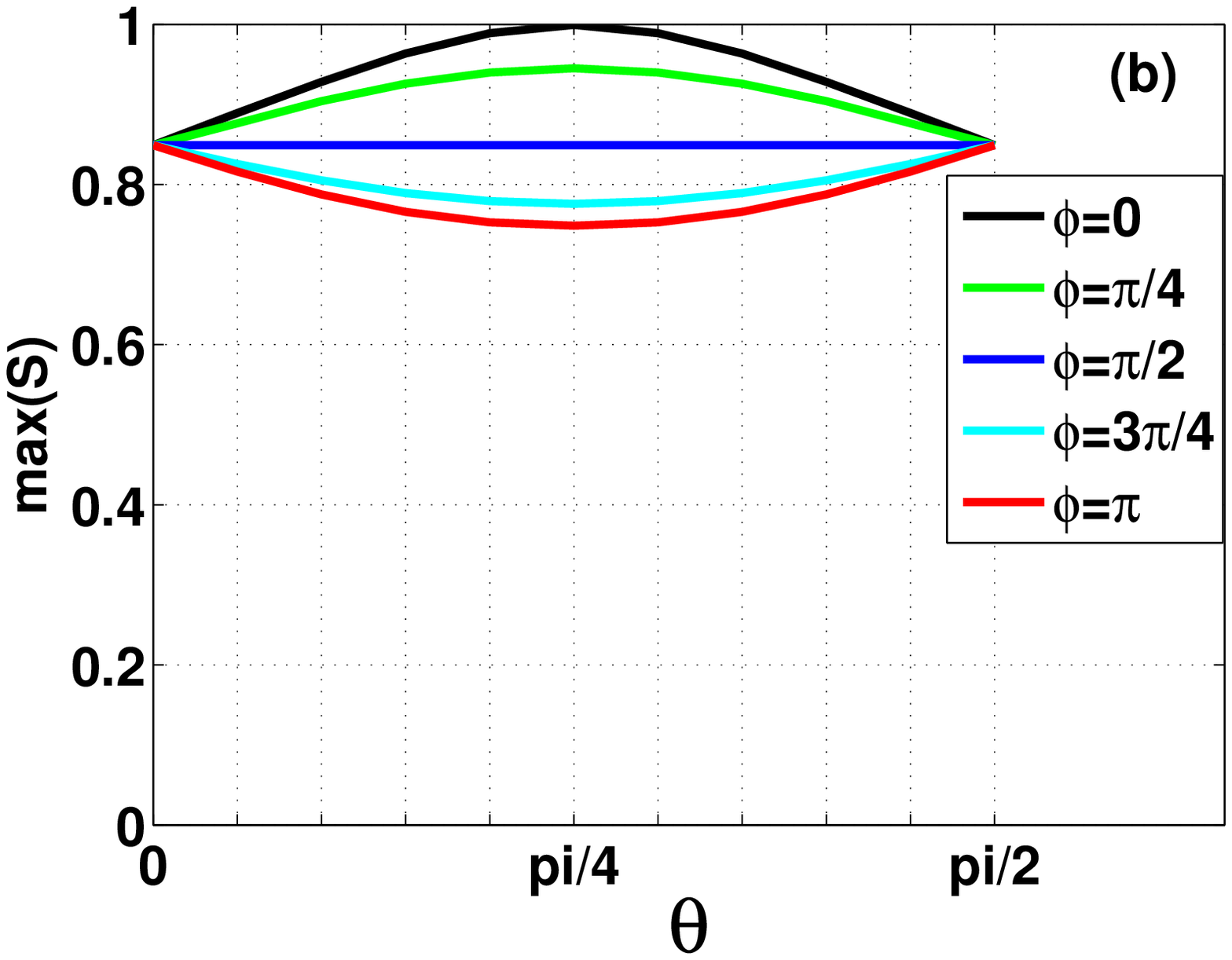}
\caption{Plot shows the maximum achievable value of delocalization as a function of $\theta$ for various values of $\phi$ for (a) $N=2$ and (b) $N=8$. We set $J=0.1$ and $\omega_j=1$ for $j=1$ to $N$. Delocalization probability is completely dependent on initial phase. Here $r=N/2$ and $s=r+1$.}
\label{fig:Fig1}
\end{figure}

If there are two photons in the array, states of the form $\ket{1}_r\ket{1}_s$  for arbitrary $r$ and $s$ or their superpositions are the TPD states. The initial state considered here is an entangled state.  We quantify  entanglement in the initial state in terms of negativity $N_e$\cite{Zyc}.  For the state $\ket{\psi}$,  $N_e=\sin2\theta/2$, independent of the  relative phase $\phi$ in the initial state.    If entanglement in the initial state  is an indicator of the degree of delocalization achievable \cite{Tang}, then $S$ should be independent of $\phi$.  
However, $P_{n,m}$ has an explicit dependence of $\phi$ which, in turn, implies that $S$ depends on $\phi$.  The maximum achievable $S$ during time evolution has been shown as a function of $\theta$ for an array of two cavities in Fig. \ref{fig:Fig1} (a) and for $N=8$ in Fig. \ref{fig:Fig1}(b). Each curve corresponds  different values of $\phi$ in the range of $0$ to $\pi$ have been considered as indicated in the figure. It is seen that as $\phi$ increases, the maximum achievable TPD decreases.   This clearly shows that the degree of TPD or TPL during time-evolution is dependent on the entanglement as well as the relative phase in the initial state.   

\begin{figure}[h!]
\centering
\includegraphics[scale=0.35]{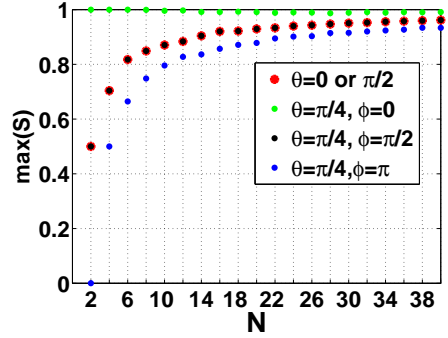}
\caption{ Maximum achievable value of S as a function of number of cavities in the array $N$ for various values of $\theta$ and $\phi$. We set $J=0.1$, $r=N/2$ and $s=r+1$. All cavities are in resonance.}
\label{fig:Fig2}
\end{figure}

  In Fig. \ref{fig:Fig2}, we have shown the maximum achievable $S$ as a function of $N$ for different choices of $\phi$ and $\theta$.   For $\phi > \pi/2$, the  maximum of $S$  increases with $N$.  If $\phi <\pi/2$, the maximum $S$ is nearly unity and remain practically at the same value.  Thus, delocalization is significant in larger arrays. \\\\\\ 
   
As an aside, we give an example of a state that never delocalizes during time evolution, 
\begin{equation}
\ket{\chi}=\frac{1}{\sqrt{N}}\sum_{n=1}^{N}(-1)^n\ket{2}_n,
\end{equation}   
where $ \ket{2}_n$ refers to 2 photons in the $n$th cavity and no photons in the other cavities.  This state is a  localized eigenstate of $H$.  Consequently,  it will never delocalize during time evolution under $H$.    

\section{Summary}  
In an array of linearly coupled cavities, two photon localization and delocalization is sensitive to both entanglement and relative phase in the initial localized state. This holds even if the initial state is maximally entangled and localized. Emergence of localization and delocalization is due to the interference of photon transfer amplitudes.  If the array has more number of cavities, the degree of delocalization increases as the constructive interference required for achieving localization is absent due to the random phases of the photon transfer amplitudes.

\end{document}